# Direct Observationof Degenerate Two-Photon Absorption and Its Saturation in WS2 and MoS2 Monolayer and Few-Layer Films

Saifeng Zhang,[1] Ningning Dong,[1] Niall McEvoy,[2,*] Maria O'Brien,[2,3] Sinéad Winters,[2,3] Nina C. Berner,[2] Chanyoung Yim,[2,3] Xiaoyan Zhang,[1] Zhanghai Chen,[4] Long Zhang,[1] Georg S. Duesberg,[2,3] and Jun Wang[1,5,*]

[1]Key Laboratory of Materials for High-Power Laser, Shanghai Institute of Optics and Fine Mechanics, Chinese Academy of Sciences, Shanghai 201800, China
[2]Centre for Research on Adaptive Nanostructures and Nanodevices (CRANN) and Advanced Materials and BioEngineering Research (AMBER) Centre, Trinity College Dublin, Ireland
[3]School of Chemistry, Trinity College Dublin, Ireland
[4]State Key Laboratory of Surface Physics, Key Laboratory of Micro and Nano Photonic Structures (Ministry of Education), Department of Physics, Fudan University, Shanghai 200433, People's Republic of China
[5]State Key Laboratory of High Field Laser Physics, Shanghai Institute of Optics and Fine Mechanics, Chinese Academy of Sciences, Shanghai 201800, China

Email: jwang@siom.ac.cn, nmcevoy@tcd.ie

## [Abstract]

The optical nonlinearity of $WS_2$, $MoS_2$ monolayer and few-layer films was investigated using the Z-scan technique with femtosecond pulses from the visible to the near infrared. The dependence of nonlinear absorption of the $WS_2$ and $MoS_2$ films on layer number and excitation wavelength was studied systematically. $WS_2$ with 1~3 layers exhibits a giant two-photon absorption (TPA) coefficient as high as $(1.0\pm0.8)\times10^4$ cm/GW. Saturation of TPA for $WS_2$ with 1~3 layers and $MoS_2$ with 25~27 layers was observed. The giant nonlinearity of $WS_2$ and $MoS_2$ is attributed to two dimensional confinement, a giant exciton effect and the band edge resonance of TPA.

Investigation of non-parametric processes concerning nonlinear absorption in transition metal dichalcogenides (TMDs) has given rise to a new category of photonic nanomaterials with potential applications in optical switching, mode-locking, optical limiting, *etc*[1-9]. Unlike gapless graphene, monolayer TMDs possess a direct bandgap. Combined with the advantage of monomolecular layer thickness, the X-M-X sandwich structure, where M stands for a transition metal (i.e., Mo, W, Ti, Nb, etc.) and X stands for a chalcogen (i.e., S, Se or Te), behaves like a natural semiconductor quantum well. Electrons are closely confined in a two-dimensional (2D) plane, implying that great optical nonlinearity enhancement is expected for such ultrathin semiconductors[10-15].

In this paper, we show that monolayer and few-layer $WS_2$ and $MoS_2$ exhibit strong two-photon absorption (TPA) for femtosecond pulses at 1030 nm. The dependence of the optical absorption nonlinearity on the layer number ($WS_2$: 1~3L, 18~20L, 39~41L; $MoS_2$: 25~27L, 72~74L) and excitation wavelength (1030 nm, 800 nm, 515 nm) was systematically investigated. The saturation of TPA for 1~3L $WS_2$ and 25~27L $MoS_2$ at 1030 nm was observed. According to a hyperbolic TPA saturation model, the TPA coefficient for 1~3L $WS_2$ at 1030 nm is deduced to be $(1.0 \pm 0.8) \times 10^4$ cm/GW, which is much larger than that of common bulk semiconductors like GaAs, GdS, ZnO, *etc*. This giant TPA coefficient is attributed to 2D confinement of electrons, a giant exciton effect and the band edge resonance of TPA.

## Results

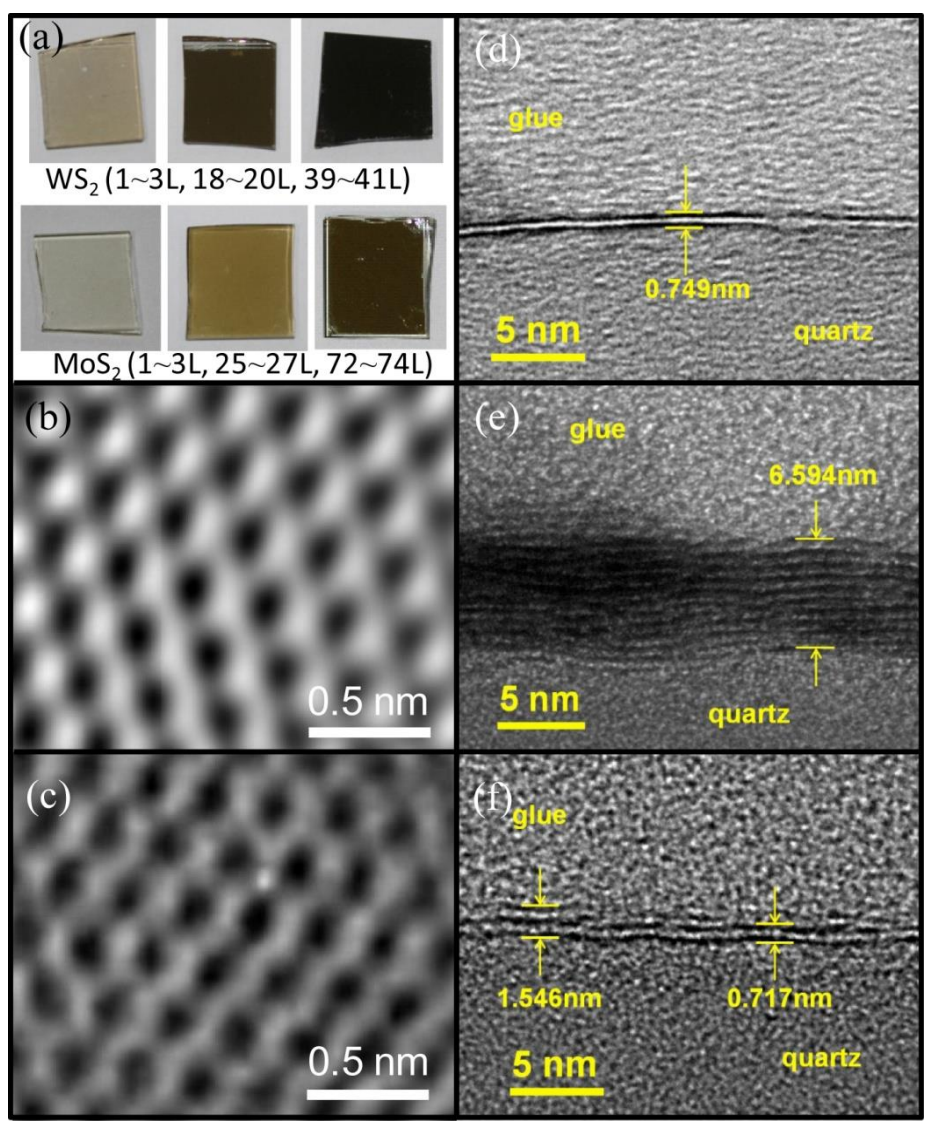

Figure 1 (a) Photographs of 1~3L, 18~20L, 39~41L WS$_2$ and 1~3L, 25~27L, 72~74L MoS$_2$; In-plane TEM of (b) 1~3L WS$_2$ and (c) 1~3L MoS$_2$; Cross-section TEM of (d) 1~3L, (e) 8~10L WS$_2$ and (f) 1~3L MoS$_2$.

The WS$_2$ and MoS$_2$ films were synthesized by direct vapor phase sulfurization of pre-deposited metal films in a quartz tube furnace with two temperature zones as reported previously[16, 17]. Thin metal films (W, Mo, 99.99% MaTecK) were sputtered onto fused quartz substrates (~10 mm×10 mm, Alfa Aesar) using a Gatan Precision Etching Coating System (PECS) with a quartz crystal microbalance to monitor the thickness of W or Mo. Four kinds of samples with initial metal film thickness of 0.5 nm, 1 nm, 5 nm and 20 nm were prepared. They were heated to 750 °C in the hot zone and annealed for 30 min under Ar flow. Then sulfur powder (MaTecK, 99%) in the upstream temperature zone was heated to its melting point (~113℃) and the sulfur vapor was transported to the pre-deposited metal films by the carrier gas (Ar). After sulfurization, the samples were held at 750 °C for 20 min and then cooled to room temperature. Photographs of $WS_2$ and $MoS_2$ films (Initial W or Mo thickness: 0.5 nm, 5 nm and 20 nm) on quartz substrates are shown in Fig. 1(a). The samples with different thickness of pre-deposited metal films are clearly distinguishable from each other. All of them appear uniform over the whole area of ~10 mm×10 mm. Fig. 1(b) and (c) show the in-plane high resolution transmission electron microscopy (HRTEM) of $WS_2$ and $MoS_2$ films. The hexagonal lattice structure in both images reveals the good crystalline quality of the samples. To determine the layer number, we utilized cross sectional TEM for the thin samples (Initial W or Mo: 0.5 nm and 1 nm) and spectroscopic ellipsometry (SE) for the relatively thick samples (Initial W or Mo: 5 nm and 20 nm). The cross-section TEM images of $WS_2$ (Initial W: 0.5 nm and 1 nm) and $MoS_2$ (Initial Mo: 0.5 nm) were obtained by plasma etching, shown in Fig. 1(d), (e) and (f). It is clear that the monolayer thickness is ~0.75 nm for $WS_2$ and ~0.72 nm for $MoS_2$, which agrees well with the reported atomic force microscopy (AFM) results[18-21]. The cross-section TEM images suggest that most part of 1~3L $WS_2$ film is monolayer, while monolayer and bilayer coexist in 1~3L $MoS_2$ film. The layer number of the thicker samples (Initial W or Mo: 5 nm and 20 nm) after sulfurization is confirmed by SE, similar to our previously reported results[22]. The initial nominal thickness of pre-deposited metal film, the thickness after sulfurization and the layer numbers are listed in Table 1.

Table 1 Layer numbers of the $WS_2$ and $MoS_2$ films

| M=W, Mo X=S | Initial nominal M (nm) | Measured $MX_2$ (nm) | $MX_2$ layer number (L) |
|---|---|---|---|
| W | 0.5 | 0.75~2.25 | 1~3 |
| | 5 | 13.5~15.0 | 18~20 |
| | 20 | 29.25~30.75 | 39~41 |
| Mo | 0.5 | 0.72~2.16 | 1~3 |

|   |   |   |
|---|---|---|
| 5 | 18~19.44 | 25~27 |
| 20 | 51.84~53.28 | 72~74 |

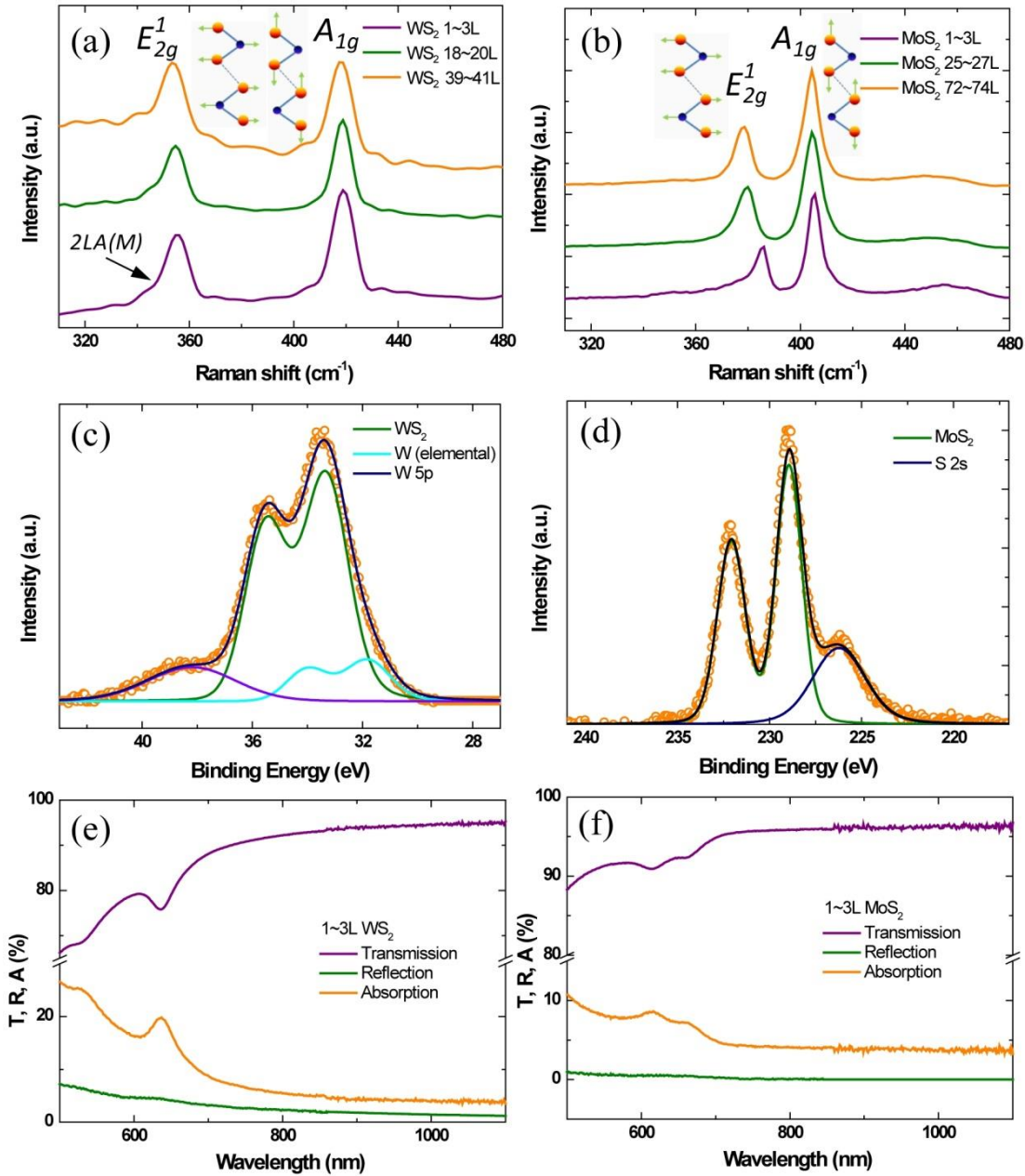

Figure 2 Raman spectra of (a) $WS_2$ and (b) $MoS_2$ films, XPS of (c) $WS_2$ and (d) $MoS_2$ films. Transmission, reflection and absorption spectra of (e) 1~3L $WS_2$ and (f) 1~3L $MoS_2$.

Raman spectroscopy was employed to examine the crystallinity and layer number. The measurements were carried out using a Renishaw inVia Raman spectrometer with a laser at 488 nm. As shown in Fig. 2(a) and (b), there are two optical phonon modes $E_{2g}^1$ and $A_{1g}$ for all WS$_2$ and MoS$_2$ films, where $E_{2g}^1$ corresponds to an in-plane optical mode and $A_{1g}$ is an out-of-plane vibration along the *c*-axis direction of the layers. Notably, the Raman peaks shift when the layer number increases. The frequency difference between $E_{2g}^1$ and $A_{1g}$ varies from 63.4 cm$^{-1}$ to 65.3 cm$^{-1}$ when the layer number of WS$_2$ increases from 1~3L to 39~41L, while it varies from 19.4 cm$^{-1}$ to 26.6 cm$^{-1}$ for MoS$_2$ from 1~3L to 72~74L. All Raman peak positions are in good agreement with reported results[18, 23-25]. The strong PL signal shown in our previous work also indicates a good uniformity of the film[17].

X-ray photoelectron spectroscopy (XPS) spectra were recorded on a VG Scientific ESCAlab MkII system using Al K_alpha X-rays and an analyzer pass energy of 20 eV. In Fig. 2(c), the deconvolution of the W 4f and W 5p$_{3/2}$ core-level region reveals the surface tungsten as predominantly in WS$_2$ and only a small component at lower binding energy which indicates traces of unsulfurized or sub-stoichiometric W. The Mo 3d core-level in Fig. 2(d) was successfully fitted with only one Mo component corresponding to MoS$_2$. Neither the W 4f nor the Mo 3d core-levels show signs of significant amounts of oxides, further indicating the high quality of the films.

In order to obtain the linear absorption coefficients of the WS$_2$ and MoS$_2$ films, we measured the reflection (*R*) and transmission (*T*) spectra in the wavelength range of 500-1100 nm with a PerkinElmer Lambda 1050 UV/Vis/NIR spectrophotometer equipped with an integrating sphere accessory. The absorption (*A*) spectrum is obtained via the formula *A=1-R-T*. In Fig. 2(e), the transmission of 1~3L WS$_2$ in the near infrared range is above 90% and it declines sharply in the visible range, while the reflection increases gradually from 1.2% at 1100 nm to 7.0% at 500 nm. The absorption in the near infrared range far from the exciton resonance peak is about 4%, nearly two times larger than that of graphene (~2.3%). The A (636.4 nm, 1.95 eV) and B (528.2 nm, 2.35 eV) exciton peaks of 1~3L WS$_2$ can be seen from the transmission and absorption curves. In Fig. 2(f), compared to WS$_2$, the transmission and reflection spectra of 1~3L MoS$_2$ exhibit relatively smooth features, above 90% for T in the whole range and about

0.04% at 1100 nm to 0.8% at 500 nm for R. The absorption in the near infrared range is about 3.8%, slightly smaller than that of $WS_2$. The exciton peaks of A (662.5 nm, 1.87 eV) and B (615.0 nm, 2.02 eV) of 1~3L $MoS_2$ can also be found in the transmission and absorption curves.

## Discussion

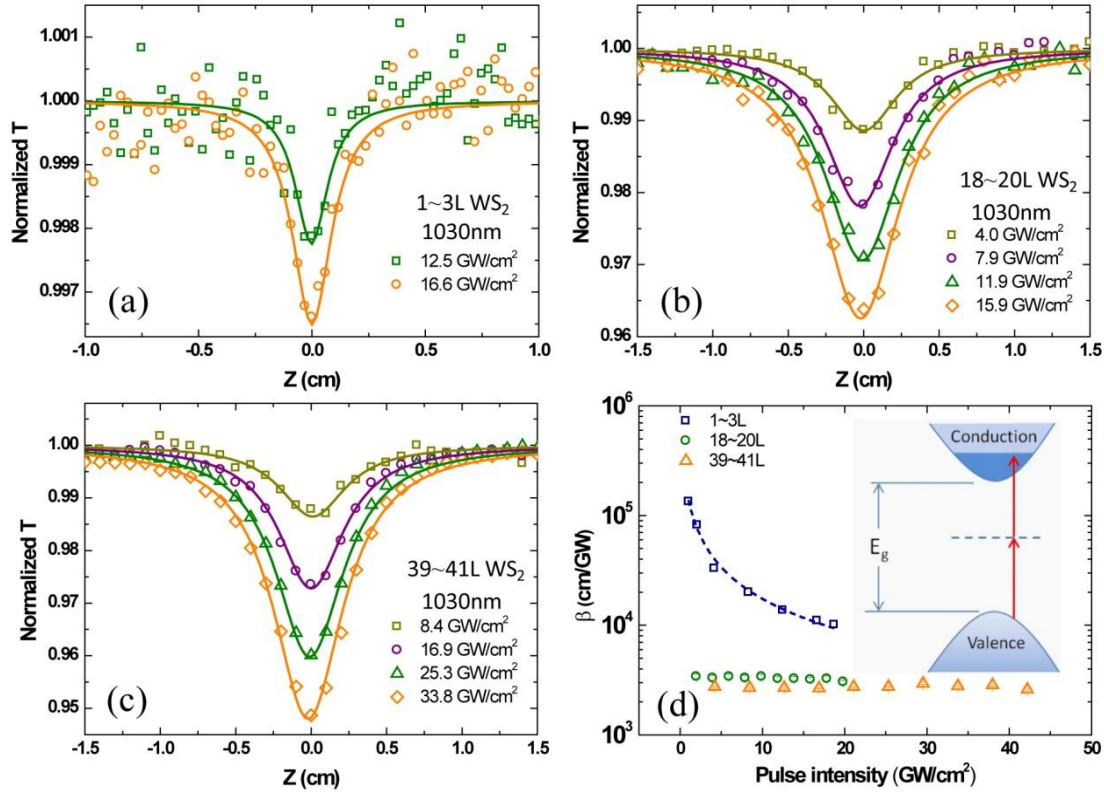

Figure 3 Z-scan results of $WS_2$ (a) 1~3L, (b) 18~20L, (c) 39~41L and (d) the corresponding TPA coefficients. Inset: the schematic of TPA saturation. (Excitation laser: 1030nm, 340fs)

The nonlinear optical (NLO) properties of the $WS_2$ and $MoS_2$ films were investigated by an open aperture Z-scan system with femtosecond laser pulses at 1030 nm, 800 nm and 515 nm. The pulse width of a fiber laser source at 1030 nm (double frequency at 515 nm) is ~340 fs with a repetition rate of 100 Hz, while the excitation laser at 800 nm is a Ti-sapphire type (pulse width: 40 fs, repetition rate: 1000 Hz, Coherent Co.). The reference beam of incident light and transmitted beam are monitored by two photodiodes as the sample moves through the focus of a lens along the laser propagation direction. Figure 3(a-c) shows the Z-scan data of 1~3L, 18~20L, 39~41L of $WS_2$ excited by 1030 nm (340 fs) laser pulses at different intensities. The NLO response of 1~3L $WS_2$ arises at 1.04 GW/cm$^2$ and the amplitude becomes the largest (~0.35%) at ~19 GW/cm$^2$. The damage threshold at 1030 nm is about 34 GW/cm$^2$. The 1~3L $WS_2$ film will be burnt through if the incident intensity is beyond this threshold and the Z-scan curve will go up abruptly. Figure 3(a) shows the typical Z-scan curves of 1~3L $WS_2$ at 1030 nm. It is reasonable to assume that TPA occurs in the 1~3L $WS_2$ film as its bandgap is about 1.9 eV[21], which is larger than one photon energy (1.204 eV) but smaller than the energy of two photons (2.408 eV) of the laser beam at 1030 nm. This is confirmed by the theoretical fitting with the nonlinear absorption model in the following part. As shown in Fig. 3(b) and (c), the 18~20L, 39~41L $WS_2$ also exhibit TPA response at 1030 nm as the bandgap decreases to about 1.3 eV when the layer number of $WS_2$ grows towards bulk[26-28]. The corresponding damage thresholds are about 20 GW/cm$^2$ and 42 GW/cm$^2$ for the 18-20L and 39-41L $WS_2$, respectively. The minimum of normalized transmission increases from ~0.35% to ~5.3% when the layer number grows from 1~3L to 39~41L, indicating the layer dependence of NLO response.

To quantitatively determine the TPA property of the $WS_2$ films, the Z-scan data were fitted by the nonlinear absorption model[29].

$$\frac{dI}{dz} = -\alpha_0 I(z) - \beta I^2(z), \qquad (1)$$

where $I$ is the intensity of the laser beam within the sample, $z$ is the propagating distance in the sample, $\alpha_0$ is the linear absorption coefficient of the sample, $\beta$ is the second order absorption coefficient. For TPA, $\beta$ is positive, while for saturable absorption (SA) $\beta$ is negative. This equation can be solved exactly and the analytical solution of the normalized transmission has the form of

$$T(z) = \frac{I(z)}{I(z-L)} \frac{1}{T_0} = \frac{-e^{\alpha_0(z-L)} + \beta \frac{te^{\alpha_0(z-L)}}{\alpha_0 + t\beta}}{-e^{\alpha_0 z} + \beta \frac{te^{\alpha_0(z-L)}}{\alpha_0 + t\beta}} \frac{1}{T_0}, \qquad (2)$$

where $t = \frac{I_0(1-R)}{1 + \frac{(z-L)^2}{z_0^2}}$, $L$ is the sample thickness, $T_0$ is the linear transmission, $I_0$ is the incident beam intensity, $z_0$ is the beam waist at the focus, $R$ is the reflection of the sample. It is worth noting that the reflection is taken into account in the model as it cannot be neglected according to the results in Fig. 2(e) and (f). The solid curves in Fig. 3(a-c) are the fitting results based on the nonlinear absorption model and it can be seen that they agree well with the experimental data. The TPA coefficients of 1~3L, 18~20L, 39~41L WS$_2$ at 1030 nm are extracted and shown in Fig. 3(d). For 1~3L WS$_2$, the TPA coefficient is about $1.34 \times 10^5$ cm/GW at ~1.04 GW/cm$^2$ and it decreases monotonously to ~$1.06 \times 10^4$ cm/GW at ~19 GW/cm$^2$. The decrease is ascribed to the saturation of TPA and it will be discussed in detail in the following part. For 18~20L, 39~41L WS$_2$, the films turn out to be indirect semiconductors and at all intensities, the TPA coefficients keep nearly constant, about 3280 cm/GW and 2750 cm/GW, respectively. Notably, the TPA coefficient of 1~3L WS$_2$ is about one order larger than that of 18~20L, 39~41L WS$_2$, which is probably due to the transition from direct to indirect bandgap caused by interlayer interaction. Compared with many other bulk semiconductors like ZnTe, CdTe, GaAs, ZnO, the TPA coefficient of monolayer and few-layer WS$_2$ at 1030 nm is about $10^3$~$10^4$ times larger[30, 31] and it is comparable to monolayer and bilayer graphene at ~1 μm[32]. It is worth noting that the TPA coefficient of WS$_2$ films does not obey the scaling rule applicable to many semiconductors proposed by E. W. Van Stryland[30]. This is attributed to the 2D confinement of layered WS$_2$, a giant exciton effect[33, 34] and the resonance of TPA near the band edge.

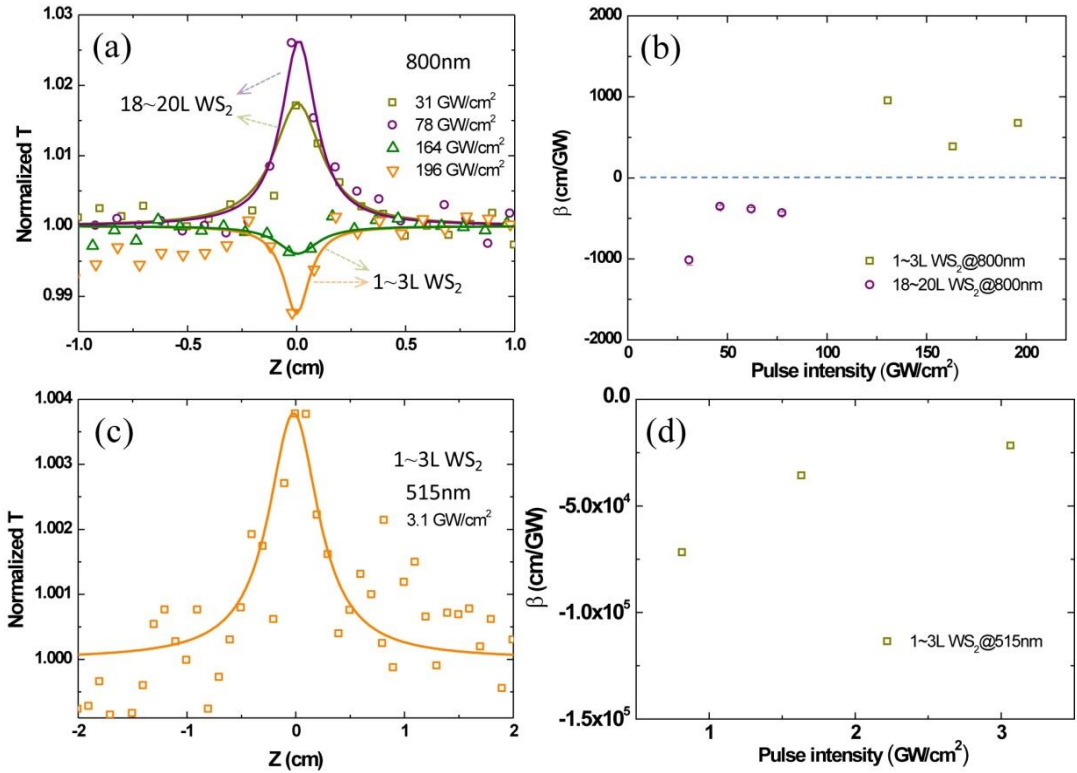

Figure 4 Z scan result of WS$_2$ (a) 1~3L, 18~20L (Excitation laser: 800nm, 40fs), (b) Nonlinear absorption coefficients at 800nm, (c) 1~3L (Excitation laser: 515nm, 340fs) and (d) SA coefficients at 515nm.

To investigate the wavelength dependence of NLO response of WS$_2$, we measured the Z-scan results at 800 nm (40 fs) and 515 nm (340 fs). As shown in Fig. 4(a), the NLO responses of 1~3L, 18~20L WS$_2$ at 800 nm (1.55 eV) are obviously different. The 1~3L WS$_2$ keeps the same TPA behavior as it does at 1030 nm, but the 18~20L film switches into SA response. The TPA excitation intensity for 1~3L WS$_2$ at 800 nm is larger than that at 1030 nm by about one order of magnitude. It may be due to the larger shift from TPA resonance of the band edge at 800 nm. The SA behavior of 18~20L sample implies that its bandgap is smaller than 1.55 eV and the SA induced by one-photon absorption takes place. The damage thresholds of 1~3L and 18~20L WS$_2$ at 800 nm are close to 196 GW/cm$^2$ and 78 GW/cm$^2$, respectively. The experimental data are fitted by the nonlinear absorption model mentioned above. The obtained TPA and SA coefficients are shown in Fig. 4(b), implying the weak saturation effect similar to the 1~3L WS$_2$ at 1030 nm. In Fig. 4(c), the 1~3L WS$_2$ at 515 nm (2.408 eV) changes into

SA behavior at a much lower intensity, which is caused by one-photon absorption as well.

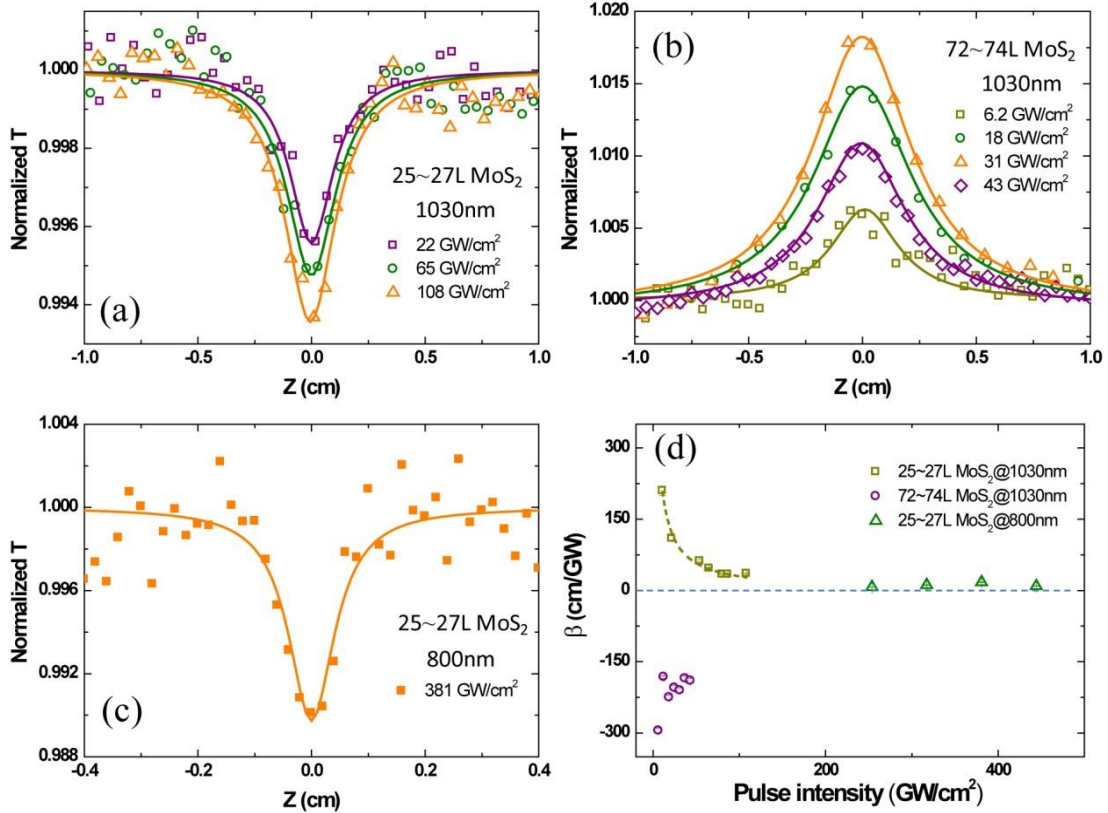

Figure 5 Z scan result of $MoS_2$ (a) 25~27L, (b) 72~74L (Excitation laser: 1030nm, 340fs), (c) 25~27L (Excitation laser: 800nm, 40fs) and (d) the corresponding nonlinear absorption coefficients.

Fig. 5 shows the layer number and excitation wavelength dependence of NLO response for 25~27L, 72~74L $MoS_2$ at 1030 nm and 800 nm. Similar to $WS_2$, the bandgap of $MoS_2$ also decreases with the increase of the layer number, from 1.90 eV for the monolayer to 1.29 eV for the bulk[35, 36]. At 1030 nm in Fig. 5(a) and (b), the 25~27L $MoS_2$ shows TPA behavior, whereas the 72~74L one changes into SA at a lower incident intensity induced by one-photon absorption. It means that the bandgap of 25~27L $MoS_2$ is larger than 1.55 eV, while that of the 72~74L one is smaller. The damage thresholds of 25~27L and 72~74L $MoS_2$ at 1030 nm are close to ~180 $GW/cm^2$ and ~43 $GW/cm^2$, respectively. At 800 nm in Fig. 5(c), the 25~27L $MoS_2$ still exhibits the TPA behavior, but the excitation intensity is much larger than that at 1030 nm in Fig.

5(a). This phenomenon also reveals that the ON/OFF resonance of TPA near the band edge plays an important role in the NLO response. The damage threshold of the 25~27L MoS$_2$ at 800 nm is close to ~444 GW/cm$^2$. The TPA coefficients of the 25~27L MoS$_2$ at 1030 nm and 800 nm are shown in Fig. 5(d). At 1030 nm, it decreases monotonously from 210 cm/GW to 34.7 cm/GW due to TPA saturation. The TPA coefficient at 1030 nm is larger than that at 800 nm for the 25~27L MoS$_2$, which keeps nearly a constant of ~11.4 cm/GW at 800 nm, confirming the importance of resonance for NLO response. The SA coefficient of the 72~74L MoS$_2$ at 1030 nm is about -250±50 cm/GW, as shown in Fig. 5(d).

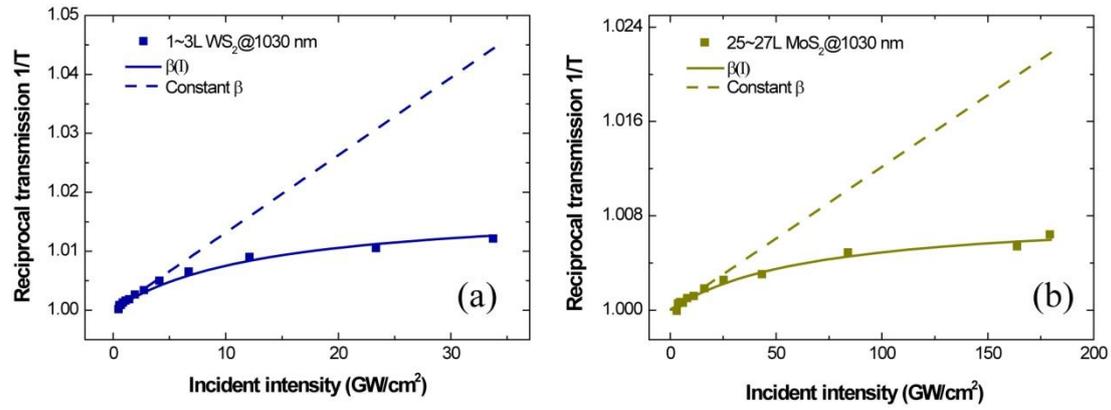

Figure 6 Reciprocal transmission versus incident intensity of (a) 1~3L WS$_2$ and (b) 25~27L MoS$_2$. Solid squares are a plot of the measured data. Solid line: theoretical variation for a hyperbolic intensity dependence of the TPA coefficient. Dashed line: theoretical variation for a constant TPA coefficient.

In Eq. (1), the nonlinear absorption coefficient $\beta$ is considered to be independent of the incident intensity and that is valid only at low intensity. When the TPA saturation occurs, the nonlinear absorption coefficient is supposed to be dependent on the incident intensity. The TPA saturation of the WS$_2$ and MoS$_2$ films can be well fitted by a hyperbolic saturation model for semiconductors[37, 38]:

$$\beta(I) = \frac{\beta_0}{1 + I/I_{sat}}, \qquad (3)$$

where $\beta_0$ is the low intensity response of the material and $I_{sat}$ is the saturation intensity of TPA for which $\beta_0$ is divided by 2. With Eq. (3), we can solve Eq. (1)

numerically and get the relationship between the reciprocal transmission 1/T and incident intensity. Figure 6(a) and (b) show reciprocal transmission versus incident intensity of the 1~3L $WS_2$ and 25~27L $MoS_2$, respectively. Solid squares are a plot of the measured data. The solid line is the theoretical variation for a hyperbolic intensity dependence of the TPA coefficient whereas the dashed line is the theoretical variation for a constant TPA coefficient. From the fitting, one can find $\beta_0 = \beta(I \ll I_{sat}) = (1.0 \pm 0.8) \times 10^4$ cm $GW^{-1}$, $I_{sat} = 13$ GW $cm^{-2}$ for the 1~3L $WS_2$ and $\beta_0 = 66 \pm 4$ cm $GW^{-1}$, $I_{sat} = 65$ GW $cm^{-2}$ for the 25~27L $MoS_2$ at 1030 nm. Recently, $MoS_2$ and $WS_2$ have been demonstrated successfully as a saturable absorber for mode-locking or Q-switching in ultrafast fiber lasers over a broad wavelength range (1 μm, 1.5 μm and 2 μm). However, the mechanism of such a saturable absorber working below the bandgap is still unclear. S. Wang[5] and R. I. Woodward *et al*.[39] ascribed it to defect-state or edge-state saturable absorption. Based on our results, it is reasonable to deduce that the TPA saturation should be easier to achieve for fewer-layer films and it might be one factor for the monolayer and few-layer $WS_2$ and $MoS_2$ working as saturable absorbers below the bandgap.

Table 2 Linear and NLO parameters of few-layer $WS_2$ and $MoS_2$ measured by Z-scan.

| Laser | sample | T (%) | $\alpha_0$ ($cm^{-1}$) | NLO response | $\beta$ (cm $GW^{-1}$) | $I_{sat}$ of TPA (GW $cm^{-2}$) | $Im\chi^{(3)}$ (esu) | FOM (esu cm) | Damage threshold $I_d$ (GW $cm^{-2}$) |
|---|---|---|---|---|---|---|---|---|---|
| 1030 nm, 1 kHz, 340 fs | 1~3L $WS_2$ | 94.76 | $7.17 \times 10^5$ | TPA | $(1.0 \pm 0.8) \times 10^4$ | 13 | $(4.0 \pm 3.6) \times 10^{-8}$ | $(1.1 \pm 1.0) \times^{13}$ | ~34 |
| | 18~20L $WS_2$ | 42.64 | $5.98 \times 10^5$ | TPA | $(3.28 \pm 0.11) \times 10^3$ | N/A | $(1.29 \pm 0.04) \times 10^{-8}$ | $(2.16 \pm 0.08) \times 10^{-14}$ | ~20 |
| | 39~41L $WS_2$ | 7.65 | $8.57 \times 10^5$ | TPA | $(2.75 \pm 0.10) \times 10^3$ | N/A | $(1.10 \pm 0.04) \times 10^{-8}$ | $(1.28 \pm 0.05) \times 10^{-14}$ | ~42 |
| | 25~27L $MoS_2$ | 92.96 | $3.90 \times 10^4$ | TPA | $66 \pm 4$ | 65 | $(4.2 \pm 0.2) \times 10^{-10}$ | $(1.10 \pm 0.03) \times 10^{-14}$ | ~180 |
| | 72~74L $MoS_2$ | 37.01 | $1.89 \times 10^5$ | SA | $-250 \pm 50$ | N/A | $(-1.50 \pm 0.33) \times 10^{-9}$ | $(7.96 \pm 1.68) \times 10^{-15}$ | ~43 |
| 800 nm, 1 kHz, 40 fs | 1~3L $WS_2$ | 92.21 | $1.08 \times 10^6$ | TPA | $525 \pm 205$ | N/A | $(2.72 \pm 0.83) \times 10^{-9}$ | $(2.51 \pm 0.77) \times 10^{-15}$ | ~196 |
| | 18~20L $WS_2$ | 35.75 | $7.22 \times 10^5$ | SA | $-397 \pm 40$ | N/A | $(-1.78 \pm 0.16) \times 10^{-9}$ | $(2.47 \pm 0.23) \times 10^{-15}$ | ~78 |
| | 25~27L $MoS_2$ | 88.98 | $6.24 \times 10^4$ | TPA | $11.4 \pm 4.3$ | N/A | $(5.26 \pm 2.46) \times 10^{-11}$ | $(8.43 \pm 3.95) \times 10^{-16}$ | ~444 |
| 515 nm, 1 kHz, 340 fs | 1~3L $WS_2$ | 67.80 | $5.18 \times 10^6$ | SA | $(-2.9 \pm 1.0) \times 10^4$ | N/A | $(-8.44 \pm 3.80) \times 10^{-8}$ | $(1.63 \pm 0.73) \times 10^{-14}$ | ~3.1 |

Table 2 summarizes the linear and NLO parameters of the WS$_2$, MoS$_2$ films measured by the Z-scan technique. The imaginary part of the third-order NLO susceptibility Imχ$^{(3)}$ can be obtained by the approximation below[2]:

$$\text{Im}\chi^{(3)} = \frac{10^{-7} c \lambda n^2}{96\pi^2}\beta, \quad (4)$$

where $c$ is the speed of light, $\lambda$ is the wavelength of the incident light, $n$ is the refractive index. The figure of merit (FOM) is defined in order to eliminate the discrepancy caused by linear absorption $\alpha_0$: $FOM = |\text{Im}\chi^{(3)}/\alpha_0|$. Imχ$^{(3)}$ of the 1~3L WS$_2$ at 1030 nm is (4.0±3.6)×10$^{-8}$ esu, about 3 times larger than the 18~20L and 39~41L ones. Imχ$^{(3)}$ of the 1~3L WS$_2$ at 800 nm is one order smaller than that at 1030 nm, indicating the OFF resonance of TPA near the band edge. Imχ$^{(3)}$ changes from positive to negative for the 18~20L WS$_2$ due to the transition from TPA to SA. The FOM also shows similar behavior. For the 25~27L MoS$_2$ at 1030 nm, Imχ$^{(3)}$ is (4.2±0.2)×10$^{-10}$ esu, about one order larger than that at 800 nm, which is also ascribed to band edge TPA resonance. For the 72~74L MoS$_2$, it changes into negative at 1030 nm due to SA.

In conclusion, we investigated the optical nonlinearity of monolayer and few-layer WS$_2$ and MoS$_2$ films using the Z-scan technique at 1030 nm, 800 nm and 515 nm with femtosecond pulses. The 2D 1~3L WS$_2$ exhibits giant optical nonlinearities having a TPA coefficient of (1.0±0.8)×10$^4$ cm/GW. The layer number and excitation wavelength dependences are systematically studied. The saturation of TPA for 1~3L WS$_2$ and 25~27L MoS$_2$ is observed. The giant nonlinearity of WS$_2$ and MoS$_2$ is ascribed to 2D confinement, giant exciton effect and the band edge resonance of TPA. The damage thresholds of the WS$_2$ and MoS$_2$ films are also given to support potential device application in the future.

## Methods:

Spectroscopic Ellipsometry

Spectroscopic ellipsometry (SE) was employed to measure the thickness of MoS$_2$ and WS$_2$[22, 40]. An Alpha SE tool (J. A. Woollam Co., Inc.) was used and SE data were obtained in the wavelength range of 380 – 900 nm at an angle of incidence of 65 ° and 70 ° with a beam spot size of ~40 mm$^2$. The SE spectra were analyzed using vender-supplied software, CompleteEASE 4.72 (J. A. Woollam Co., Inc.).

The measured SE spectra consist of psi (Ψ) and delta (Δ) components which represent the amplitude ratio (Ψ) and phase difference (Δ) between p- and s-polarizations, respectively. The two parameters are related to the ratio ρ, defined by the equation of $\rho = r_p/r_s = \tan(\Psi)\exp(i\Delta)$, where $r_p$ and $r_s$ are the amplitude reflection coefficients for the p-polarized and s-polarized light, respectively[41].

Optical models built for SE analysis have a four-layered structure, where there are three sub-layers (Si substrate, an interface layer between Si and $SiO_2$, a $SiO_2$ layer) and a top material layer ($MoS_2$ or $WS_2$). Tauc-Lorentz (T-L) oscillation model was used to determine the thicknesses of the $MoS_2$/$WS_2$ thin films[42].

## Acknowledgements

The authors want to thank the support of NSFC funds (No. 61308034, 61178007), STCSM (No. 12ZR1451800) and the External Cooperation Program of BIC, CAS (No. 181231KYSB20130007). NME acknowledges SFI for 14/TIDA/2329. MO, SW, NCB, CY and GSD acknowledge SFI for PI_10/IN.1/I3030.